\documentstyle[draft,epsf]{mn}

\title{The bimodal spiral galaxy surface brightness distribution}

\author[Bell \& de Blok  ]
{
Eric F. Bell$^1$ and W. J. G. de Blok$^2$
\thanks{Bolton Fellow}
\thanks{Present address: Australia Telescope National Facility, 
	PO Box 76, 
	Epping NSW 1710, 
	Australia} 
\\
$^1$ Department of Physics, University of Durham, Science 
Laboratories, South Road, Durham DH1 3LE, UK \\
$^2$ Astrophysics Group, School of Physics, 
           University of Melbourne, 
             Parkville, VIC 3052, 
                 Australia}

\begin{document}
\date{Submitted to MNRAS: \today}

\maketitle

\begin{abstract}
We have assessed the significance of Tully \& Verheijen's
(1997) bimodal Ursa Major Cluster spiral galaxy near-infrared 
surface brightness distribution, 
focussing on whether this bimodality is simply an 
artifact of small number statistics.   
A Kolmogorov-Smirnov style of significance test shows that the
total distribution is fairly represented by a single-peaked distribution,
but that their isolated galaxy subsample (with no 
significant neighbours within a projected distance of 
$\sim$ 80 kpc)
is bimodal at the 96 per cent level. 
We have also investigated the assumptions underlying
the isolated galaxy surface brightness distribution, 
finding that the (often large) inclination
corrections used in the construction of this distribution
reduce the significance of the bimodality.
We conclude that the 
Ursa Major Cluster dataset is insufficient to 
establish the presence of a bimodal near-infrared surface 
brightness distribution:  an independent sample of $\sim$ 100 isolated, low
inclination galaxies is 
required to establish bimodality at the 99 per cent level.
\end{abstract}


\section{Introduction}

The distribution of spiral galaxy disc surface brightnesses
is of great observational and theoretical importance in
extragalactic astronomy.  Observationally, knowledge about the present day
disc galaxy central
surface brightness distribution (SBD) allows us to better understand the
importance and implications of surface brightness selection effects,
which is important for e.g. understanding the local galaxy luminosity
function (e.g.\ Dalcanton 1998; de Jong \& Lacey 1998), 
or in comparing galaxy populations at low and high redshifts
(e.g.\ Simard et al.\ 1999).
Theories of galaxy formation and evolution link the SBD
with the angular momentum, mass and star formation
history of galaxies:  indeed, the SBD has
been used as a constraint on some galaxy evolution models 
\cite{dalc97,mo1998,baugh1998}.

Previous studies have generally found a SBD that can be described as
flat or slowly declining with surface brightness, modulated by a sharp
cut-off at the high surface brightness end, normally associated with
the Freeman \shortcite{f70} value (e.g. de Jong 1996c; McGaugh 1996).

Tully and Verheijen (1997; TV hereafter) derived the SBD of a complete
sample of 60 spiral discs in the Ursa Major Cluster from an imaging
study in the optical and near-infrared \cite{tv1}. Their data
suggested an inclination-corrected near-infrared 
{\it bimodal} SBD (as opposed to
the conventional {\it unimodal} distribution described above) with a
peak associated with the Freeman value, and a second peak $\sim 3$ mag
arcsec$^{-2}$ fainter, separated by a gap (Fig.\ 1, solid line in panel a).  
This would have far-reaching implications for our understanding of galaxy
formation and evolution.  A bimodal SBD would imply that there are two
discrete galaxy populations: a high surface density population and one with
surface densities a factor of $\sim 10$ lower (assuming comparable
mass to light ratios for the two populations).  Correspondingly, the
masses, spin distributions and star formation histories of these two
families of galaxies would adopt one of two preferred states, and
plausible mechanisms for generating these discrete differences in
e.g.\ star formation history, or spin distribution, would have to be
found.

Why this bimodal distribution was not found before is an important
question.  TV address this problem at length: here we will summarise
their main arguments.  First, the bimodal SBD
is based on near-infrared $K'$ data and, in fact, is {\it
only} really apparent there.  Near-infrared photometry is much
less susceptible to the effects of dust extinction than optical
photometry.  Furthermore, high surface brightness galaxies are 
typically redder than low surface brightness galaxies (de Jong 1996b),
which accentuates the gap in the near-infrared.  
TV demonstrate that in the optical $B$ band the bimodality
is washed out by these effects:  the bimodality in the optical is 
consistent with the bimodality in the near-infrared, but is less significant
(accordingly, we do not consider TV's
optical SBDs here).  A bimodal SBD would therefore remain undetected in optical
studies (e.g. McGaugh 1996, de Jong \& Lacey 1998).

Second, the TV SBD has  been determined from
a complete sample of galaxies at a common distance,
making selection effects much easier to understand.  This may explain 
why de Jong \shortcite{dej}, using $K$ band photometry for a sample of field
spiral galaxies, did not report any sign of bimodality in his SBD.

In an attempt to investigate whether the bimodal distribution could be
caused by environmental differences, TV defined an isolated
subsample, consisting of 36 galaxies that have a projected distance
$>80$ kpc to their nearest neighbour 
(assuming a distance to the Ursa Major Cluster of 15.5 Mpc, 
corresponding to $H_0 = 85$ kms$^{-1}$\,Mpc$^{-1}$). 
They found that the gap is
more pronounced in this isolated subsample, and attributed
intermediate surface brightnesses found in their total sample to
interactions (Fig.\ 1, solid line in panel b).

Unfortunately, TV did not attempt to robustly estimate the
significance of the apparent bimodality.
Their only estimate is based on the isolated subsample where
they compare the number of galaxies in the 1.5 mag wide gap 
(between $\sim$ 17.5 and $\sim$ 19 mag\,arcsec$^{-2}$; 3 galaxies) with
the number of galaxies that they expected to see in that gap (20 galaxies).
This expectation value was derived from a model unimodal distribution
normalised to reproduce the peak number of galaxies in the high
surface brightness bins, {\it not} the total number of galaxies.
Properly normalising the model distribution lowers the expectation 
from $\sim$ 20 galaxies to $\sim 12$ galaxies.

In this paper, we investigate the significance of the bimodal
near-infrared SBD, focussing on whether this bimodality is simply an 
artifact of small number statistics.  In sections 2 and 3, we 
use a Kolmogorov-Smirnov (KS) style of test
to address this question, investigating
the significance of both the total and isolated samples.
In section 4, we discuss the results, focussing on the r\^{o}le
of inclination corrections.  
We present our main conclusions in section 5.

\section{Constructing a null hypothesis}

\begin{figure}
\begin{center}
  \leavevmode
  \epsffile{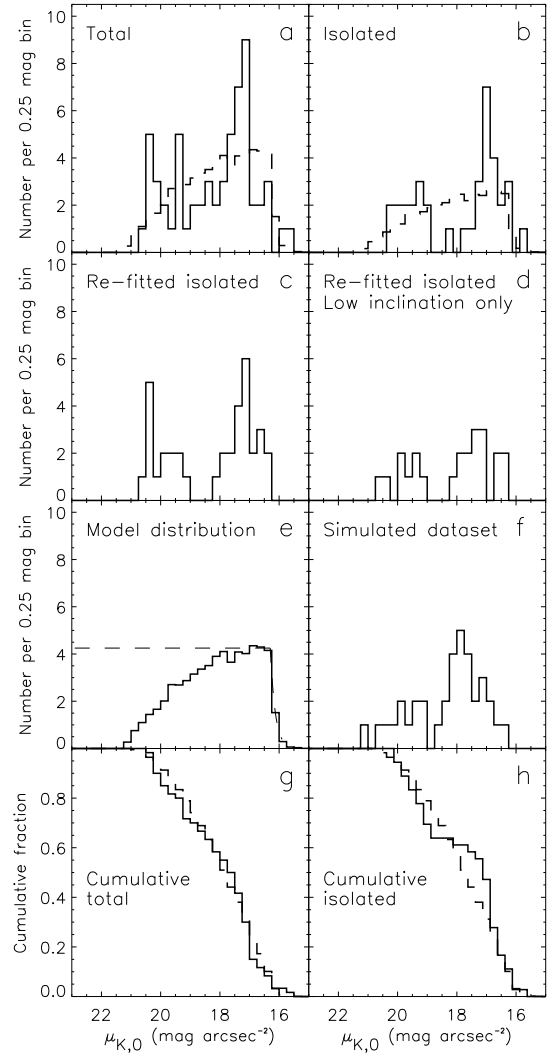}
\end{center}
\vspace{-0.3cm}
\caption{
The SBD for the total sample
of 60 galaxies (solid line in panel a) 
and for the isolated subsample of 36 galaxies (solid line in panel b).
Also shown in panels a and b is a normalised representative
null distribution (dashed line).
Panel c contains our re-fitted isolated galaxy subsample, 
and panel d shows the low inclination isolated galaxies only.
Panel e contains a representative model distribution
taking selection effects into account (solid line), and
shows the same distribution without taking into account the selection
effects (dashed line, with an arbitrary normalisation).
Panel f shows a 36 galaxy realisation of the
single-peaked model distribution, chosen to 
appear reasonably similar to the isolated
galaxy subsample.  Panels g and h show the cumulative 
distribution function for the total and isolated subsamples
respectively.  Solid lines depict the observed cumulative
SBDs and dashed lines represent the
model distribution.
}
\label{fig:diff}
\end{figure}

In order to test the significance of the bimodality in the SBD, it
is necessary to choose a null distribution to test the observations against.
Ordinarily, it would be best to choose some kind of `best bet'
observationally determined or theoretically motivated 
null hypothesis:  failure of this null hypothesis would indicate
that it does not accurately describe the dataset.  However, 
in this case, we wish to quantify the significance of a particular
feature in the observational distribution, namely the bimodality 
in the Ursa Major Cluster SBD.  For this reason, it is necessary to 
be careful about choosing a null distribution: if a null distribution
was chosen which did not adequately describe the general features of 
the SBD, then a high significance (i.e.\ a low probability of obtaining
the data from the null) would {\it not} represent a large degree of
bimodality, but rather would indicate that the null hypothesis was a 
poor match to the general shape of the SBD.

For this reason, we have chosen to fit a relatively simple model
distribution to the data (taking into account selection 
effects), allowing us to determine
the significance of the bimodality by minimising the mismatches
between the general shapes of the two distributions.

For the unimodal model, we use a slightly modified version of the optical 
SBD presented in McGaugh \shortcite{mcgaugh}: 
\begin{eqnarray}
\log_{10}\Phi(\mu_{K',0}) & =  \nonumber
m (\mu_{K',0} -  \mu_{K',0}^*)\hspace{2.5cm} \\
 \hfill{\rm where}&
\left\{
\begin{array}{llll}
m &= &m_f  &{\rm if\ }  \mu_{K',0} \ge \mu_{K',0}^* \\
m &= &2.6  &{\rm if\ }  \mu_{K',0} < \mu_{K',0}^*
\end{array} \nonumber
\right.
\end{eqnarray}
The critical surface
brightness $\mu_{K',0}^*$ and the faint-end slope $m_f$ are adjustable
parameters which are fit below.
\footnote{McGaugh's original values were $\mu_{0}^* = $ 21.5 $B_J$
mag\,arcsec$^{-2}$ and $m_f =-0.3 $, which are roughly consistent 
with the best fit results described in section 3.}
For reasonable values of $m_f$, the number of 
galaxies per unit mag\,arcsec$^{-2}$ only slowly changes
with surface brightness, and falls quite sharply to zero above a given
`critical' surface brightness $\mu_{K',0}^*$ (see e.g.\ the dashed
line in panel e of Fig.\ \ref{fig:diff}). 

To take into account the selection effects that affect TV's
observations, it is also necessary to adopt a scale size distribution:
we assume that the scale size distribution is constant (per
logarithmic interval) over the interval $\log_{10} h_{K'} = 0.8$ to
$\log_{10} h_{K'} = 1.7$ ($\sim$6.3 arcsec to 50.1 arcsec, or, at an
adopted distance of 15.5 Mpc, $\sim$0.5 to 3.8 kpc).

Selection criteria are modelled by choosing only galaxies with
isophotal magnitudes $m_{K'} < 12$, and isophotal diameters $D_{K'} > 60''$,
both measured at the 22.5 $K'$ mag\,arcsec$^{-2}$ isophote.  Though
these criteria differ slightly from those used by TV (as their
selection was made in $B$ band rather than $K'$ band), they are
comparable to better than 0.5 mag, assuming $B-K' \sim 3$ for faint
late-type galaxies \cite{dej2}, and accurately follow the surface
brightness and size limits on TV's $\mu_{K',0} - \log_{10}h_{K'}$ plane.
We emphasize that we do not hope to somehow represent the 
`universal' $K'$ band SBD with this function, we simply aim to 
construct a plausible, simple SBD that provides a sensible
single-peaked fit to the observed bimodal Ursa Major SBD.

In panel e in Fig.\ \ref{fig:diff}, we show a 10$^4$ galaxy 
Monte Carlo realisation 
of a model distribution with $\mu_{K',0}^* = 16.3$ and $m_f = 0.0$ 
(also in panels a and b, and in  
cumulative form in panels g and
h as a dashed line). 
We also show the same model distribution without the selection effects 
(with arbitrary normalisation) in
panel e as a dashed line.

\section{Results}

In the above section, we described the single-peaked 
distribution that we will fit to the observed SBD.
As we want to take into account selection effects in the null
distribution, we cannot make a straightforward fit, but have to use
Monte Carlo realisations of the null distribution and associated 
selection effects.  We generate a grid of Monte Carlo 
unimodal distributions, each containing 10$^4$ galaxies
(later renormalised to contain the same number of galaxies in the
observed SBD), and
each with different values of $\mu_{K',0}^*$ and $m_f$ (with 
grid steps of 0.05 in $\mu_{K',0}^*$ and 0.02 in $m_f$).
The best fit unimodal
distribution is determined by minimising the 
Cash statistic $C = -2 \sum_{i=1}^{n_{bins}} \log
P_i(n,\lambda)$, where $P_i(n,\lambda)$ is the Poissonian probability
of observing $n$ galaxies in a bin which the unimodal 
model predicts should have
a mean galaxy number of $\lambda$ \cite{cash}. 
The Cash statistic has the advantage
that the underlying distribution does not have to be Gaussian.  
As the model grid used to determine the best fit is 
itself a Monte Carlo realisation of the true underlying model 
grid, the best fit parameters will depend slightly on the model
grid realisation used.  This ultimately affects the 
significances that we will derive later.  We have therefore
run the code 10 times, and thus constructed 10 realisations 
of the entire model grid, and derived a best fit for each 
of these realisations.  The best fit unimodal distribution
typically has parameters
around $\mu_{K',0}^* \sim 16.35 \pm 0.05$
mag\,arcsec$^{-2}$ and a faint end slope $m_f \sim 0.05 \pm 0.08$,
where these uncertainties are the RMS variation in the fit parameters.

To estimate the significance of the departures of the observed SBD
from the best fit model distribution we use the KS
statistic $d_{\rm max}$, defined as the maximum distance between the
cumulative distributions of the model and the data (i.e.\ the
maximum distance between the solid and dashed lines in
panels g and h of Fig.\ \ref{fig:diff}).  The $d_{\rm
max,obs}$ between this best fit model and observed SBD is then
measured.  

However, because we have {\it fitted} a single peaked distribution
to the bimodal SBD, it is not possible to convert the measured
$d_{\rm max,obs}$ into a probability in the standard way 
(e.g. Press et al.\ 1986).
\footnote{Note that the binning of the data in a histogram also,
strictly speaking, invalidates the use of a standard KS test, 
although if the binning is over a much smaller scale than
the `structure' in the SBD (as it is in this case), tests
have shown that the effects on the final significances are 
negligible.}
Therefore, it is necessary to use 
Monte Carlo simulations to convert the measured maximum distance
into a probability.  Accordingly, we 
generated 100 random simulated 
datasets from the best fit model, and subjected
them to the {\it same fitting and maximum distance measuring 
procedure as the data}.  In this way, we determined what the 
distribution of maximum distances should be from the simulations.
The fraction of simulations with maximum distances larger than
the $d_{\rm max,obs}$ then gives us the probability $P$ that
the observed dataset was a statistical accident.

Because the model unimodal distributions are Monte Carlo
realisations of the true model distributions, this procedure
(i.e.\ generating the 100 datasets and measuring the significance)
was repeated for each of the 10 model grid realisations described above,
to allow more accurate measurement of the significance.  
The significance is defined as $1 - P$, with $P$ defined as above.
Significances for
the total sample of 60 galaxies, and the isolated sample of 36
galaxies are given in Table \ref{tab:sig}.
Quoted significances are the mean (along with the error in those
means) of the 10 runs.  Note that we also carried out this analysis in
two different programming languages on two different computers using
different sets of random number generators: the results were found to
be indistinguishable to within the quoted errors.  

\begin{table}
\caption{The significance of the bimodality in the Ursa Major Cluster} 
\label{tab:sig}
\begin{center}
\begin{tabular}{lcc}
\hline
Sample & $N_{\rm gal}$ & Significance \\
\hline
total & 60 & 58\% $\pm$ 4\% \\
isolated & 36 & 96.0\% $\pm$ 0.5\% \\
isolated (re-fit) & 36 & 86\% $\pm$ 1\% \\
low $i$ isolated & 23 & 42\% $\pm$ 3\% \\
\hline
\end{tabular}
\end{center}
\end{table}

\section{Discussion}

\begin{figure}
\begin{center}
  \leavevmode
  \epsffile{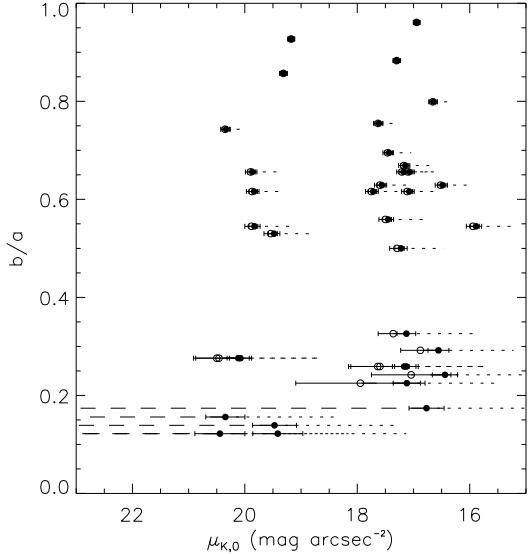}
\end{center}
\vspace{-0.3cm}
\caption{
Uncertainties in inclination corrected surface brightness against the 
axial ratio $b/a$.  We show the surface brightness of the galaxy corrected
for inclination assuming an infinitely thin disc (solid circles) connected
by a dashed line to the inclination corrected surface brightness assuming
an intrinsic disc axial ratio $q_0$ of 0.20 (open circles).  
Error bars denote the 
uncertainty in surface brightness due to (a rather generous) error
in axial ratio of 0.05.  Dotted lines connect the inclination corrected
surface brightness to the {\it uncorrected} surface brightness.
}
\label{fig:inclin}
\end{figure}

From Table \ref{tab:sig}, it is clear that the bimodality shown by
the total sample is not statistically significant.  
However, the isolated subsample shows a bimodality significant 
at the 96 per cent level.
To understand if this high significance of the histogram for the 
isolated subsample reflects a true, astrophysical bimodality in 
the Ursa Major Cluster, it is necessary to understand the 
assumptions used to construct the isolated galaxy SBD.

An obvious check is to see whether systematic effects in the data can
produce a bimodality, e.g., because of bulge contamination in the
fits.  We re-fit the surface brightness profiles presented in Tully et
al.\ \shortcite{tv1} using the `marking the disc' technique (the 
same technique that was used by TV), 
to check that there were no obvious personal biases in the 
fitting of the surface brightness profiles.
Comparing our $K'$ band central surface brightnesses with those of
Tully et al.\ \shortcite{tv1}, we found a mean offset of +0.16
mag\,arcsec$^{-2}$, and that 68 per cent of the central
surface brightnesses compared to better than 
$\pm$ 0.25 mag\,arcsec$^{-2}$ (both with and without 
this mean offset applied).
When corrected for 
inclination in exactly the same way as TV
(assuming an infinitely thin disc; by 
subtracting $2.5 \log_{10}(b/a)$, where $b/a$ is the
minor to major axis ratio), the two versions of the isolated subsample
SBD are virtually indistinguishable (see Fig. \ref{fig:diff}, panel
c), and the significance of the re-fitted distribution is hardly
different, decreasing slightly to 86 per cent.  
This suggests that
there are no systematic personal biases in the fits 
TV use that might introduce a spurious bimodality.
Note that de Jong \shortcite{dejong1996ii} shows that the 
`marking the disc' technique is unbiased, with $\sim$ 90 per cent of the 
central surface brightnesses accurate to 0.5 mag, 
compared to more sophisticated bulge-disc decomposition techniques.

However, because there was no inclination cut applied to
the Ursa Major sample, 
some of the inclination corrections that were applied to the
$K'$ band surface brightnesses were rather large: in $\sim$ 1/3 
of the cases (in 
either the total or isolated samples) 
they exceeded 1 mag.  However,
large inclination corrections are highly uncertain for a number of
reasons, even in $K'$ band.  

In Fig.\ \ref{fig:inclin}, we show only two of the many
uncertainties that affect inclination 
corrections in $K'$ band.  
First we computed the corrected surface
brightnesses, assuming i) an infinitely thin disc (solid circles) and
ii) an intrinsic disc axial ratio $q_0 = 0.2$ (Holmberg 1958; 
open circles).  These two
values are connected with a dashed line in Fig.\ \ref{fig:inclin}. 
The error bars in Fig.\ \ref{fig:inclin} represent the random
uncertainty in the inclination correction due to axial
ratio errors of 0.05.
Taken together, this 
shows that the inclination corrections for galaxies 
with axial ratios $b/a$ below 0.4 have considerable systematic uncertainties
stemming from our poor knowledge of the intrinsic axial ratios of galaxies, 
coupled with the random uncertainty associated with measurement error.
Taken together, consideration of these two uncertainties argues that 
the inclination corrections applied in this paper and by TV 
are likely to be lower limits.

However, a number of effects conspire to make inclination corrections
smaller, suggesting that the inclination corrections might actually be 
upper limits.  Firstly, the effects of dust in nearly
edge-on galaxies is non-negligible, even in the $K'$ band, 
which could lower the inclination correction by roughly as much as 
0.3 mag \cite{global}.   Secondly, averaging of the surface
brightness profiles over elliptical annuli 
at high inclinations (as in this case) 
is likely to produce systematic underestimates of the real central surface 
brightness \cite{hu94}.  Finally, at large inclinations, the assumption of a
thin, uniform slab disc breaks down:  vertical and radial 
structure in the galaxy will affect the inclination correction in 
non-trivial ways.  Accordingly, in Fig.\ \ref{fig:inclin} 
we have also connected the 
inclination corrected surface brightness with the uncorrected surface
brightness with a dotted line.

In Fig.\ \ref{fig:inclin}, we can see that 
uncertainties in these large inclination corrections may go some way
towards filling in the gap in the SBD. 
As an example, we consider the four galaxies with the largest
inclination corrections.
UGC 6667, UGC 6894,
NGC 4010 and NGC 4183 have $> 2$ mag inclination corrections and uncorrected
(reasonably high) $K'$ surface brightnesses between 17 and 18.5
mag\,arcsec$^{-2}$. The large corrections move them across the gap
from the high surface brightness into the low surface brightness regime.
However, if these corrections are at all
uncertain, they may in reality lie in the gap of
the $K'$ band SBD.  Moving even one or two of them into the gap in the
isolated galaxy SBD would decrease markedly the significance of
the bimodality in the isolated subsample.  Certainly a few of the
galaxies mentioned above suffer from problems that may seriously affect the
value of the inclination correction (not because of the uncertainty 
in axial ratio, but because of uncertainties in e.g.\ the vertical and
radial structure in the galaxy): NGC 4010 is clearly lopsided, with one
half of the galaxy puffed up with respect to the other half, while NGC
4183 has a clear warp.

Because of these serious uncertainties in the larger inclination
corrections, and because there are no clear recipes for producing
more realistic inclination corrections which take into account dust
and the vertical and radial structure of the disc, the only
fair thing to do is omit the high-inclination galaxies from the SBD.
Accordingly, the high-inclination galaxies were removed from the
isolated subsample, leaving 23 galaxies with $b/a$ larger than 0.4,
corresponding to an inclination of less than 66$^{\circ}$ or
inclination corrections smaller than 1 mag.  The resulting SBD is
shown in Fig.\ \ref{fig:diff}, panel d:  the
bimodality is insignificant in this SBD, due primarily to small number 
statistics (Table \ref{tab:sig}).
To check how large the sample size needs to be to
detect significant bimodality in the low inclination isolated
subsample, we simulated intrinsically bimodal datasets and tested them
using the above procedure.  Bimodal datasets were generated using two
normal distributions with $\sigma = 0.5$ mag and means of 17.15 and
19.85 mag\,arcsec$^{-2}$, with the high surface brightness peak
containing on average 63 per cent of the galaxies (thus resembling the
TV distribution).  These simulations showed that in order to detect a
bimodality at the 95 (99) per cent level in the low inclination sample
at least 4/5 of the time, the sample size needs to be increased to
around 50 (100) galaxies.

\section{Conclusions}

We have re-analysed Tully \& Verheijen's (1997) 
bimodal SBD, using a 
Kolmogorov-Smirnov style of test to estimate the likelihood
that a single-peaked distribution would be able to 
reproduce the bimodality in the Ursa Major Cluster.
\begin{itemize}
\item The total sample of 60 galaxies 
is inconsistent with a single-peaked distribution at only the 58 per
cent level.
\item However, the isolated subsample 
of 36 galaxies is significantly bimodal: it
is inconsistent with the null distribution at the 96 per
cent level.
\item We re-fit the $K'$ band surface brightness profiles of the
Ursa Major sample, and found that this re-fit made relatively 
little difference, with the isolated subsample retaining a 
reasonable significance of 86 per cent.
\item However, we argue that large inclination corrections are
uncertain even in the near-infrared, placing the reality of the gap 
between high and low surface brightness galaxies in some doubt.
\item When the galaxies with inclination corrections $> 1$ mag
are removed from the isolated subsample, the significance of the
23 galaxy subsample drops to only 40 per cent.  
\item Assuming that the low inclination, isolated galaxy SBD
is truly bimodal, in order to increase the significance
to 95 (99) per cent, it is necessary to increase the low inclination 
isolated galaxy sample size by
around a factor of two (four).
\end{itemize}
To summarise, if inclination corrections greater than 1 mag can be
trusted, then there is reasonable evidence that the Ursa Major Cluster
SBD is bimodal.  If, however, these
large inclination corrections are in some doubt, then the SBD lacks a
sufficient number of galaxies to argue convincingly against a unimodal
SBD.  Either way, to convincingly demonstrate 
at the 99 per cent level that the near-infrared spiral galaxy 
SBD is bimodal will require an independent dataset to be used
with around a factor of four more low inclination, isolated galaxies
than are included in the Ursa Major Cluster sample.

\section*{Acknowledgements}

We would like to thank Marc Verheijen for providing tables of data
in electronic form.
We would also like to thank the referee for their comments on the manuscript.
We would like to thank Greg Bothun, Roelof Bottema, Richard Bower, 
Roelof de Jong, Rob Kennicutt, John Lucey and 
Stacy McGaugh
for useful discussions and comments on the manuscript.
EFB would like to thank the Isle of Man Education Department for their
generous support.
This project made use of STARLINK facilities in Durham.

\end{document}